\documentclass[preprint,amsmath,amssymb,aps,]{revtex4-1}
\pdfoutput=1

\usepackage{graphicx}
\usepackage{color}
\usepackage{stmaryrd}

\def\QTM{QTM}
\def\e{\text{e}}
\def\i{\text{i}}

\begin{document}


\title{A Hubbard model with integrable impurity}

\author{Yahya {\"O}z}
\affiliation{%
 Turkish Aerospace, 06980 Ankara, Turkey}%
 \altaffiliation[Also at ]{University of Wuppertal}
 \email{yahya.oz@tai.com.tr}
\author{Andreas Kl{\"u}mper}%
\affiliation{%
 University of Wuppertal, Faculty of Mathematics and Natural Sciences, 42119 Wuppertal, Germany
 }%

%

\date{\today}

\begin{abstract}

We construct an integrable Hubbard model with impurity site containing spin
and charge degrees of freedom. The Bethe ansatz equations for the Hamiltonian
are derived and two alternative sets of equations for the thermodynamical
properties. For this study, the thermodynamical Bethe ansatz and the quantum
transfer matrix approach are used. The latter approach allows for a consistent
treatment by use of a finite set of non-linear integral equations.  In both
cases, TBA and QTM, the contribution of the impurity to the thermodynamical
potential is given by integral expressions.
\end{abstract}

\maketitle


\section{\label{sec:level1}Introduction}
The history of the Hubbard model as an exactly solvable model started in $1968$
with Lieb and Wu's article \cite{LW68}. The consistency of the Bethe Ansatz, i.e. absence of multi-particle processes was shown by Essler and Korepin \cite{EK94,EK94+} and transparently by the embedding of the Hubbard model into a formulation with commuting transfer matrices by the work of Shastry \cite{Sha86}. Following this, the algebraic Bethe ansatz constructions of the eigenstates were given by Martins \cite{RM97,MR98}. 

Lieb and Wu demonstrated that the Bethe
ansatz can be used which reduces the eigenvalue problem of
the Hamiltonian to solving a set of coupled algebraic equations, which are called Lieb-Wu equations. They calculated the ground state energy and
discovered that the Hubbard model undergoes a Mott metal-insulator
transition at half filling (on average one electron per site) with critical
interaction strength $U=0$.  

A classification of the solutions of the Lieb-Wu equations in terms of the
so-called string hypothesis was given by Takahashi in 1972 \cite{Tak72}. He used this to replace the
Lieb-Wu equations by simpler ones that describe the scattering states of bound
complexes (strings), and derived the thermodynamic Bethe ansatz (TBA)
equations, which determine the Gibbs free energy of the Hubbard model. From
the TBA, in the limit of low temperatures, Takahashi calculated the specific
heat \cite{Takah74}. Subsequently a rather reliable picture of the thermodynamics of
the Hubbard model was derived from numerical solutions of the thermodynamic
Bethe ansatz equations \cite{KUO89,UKO90}. As a matter of fact, Takahashi's equations, in
conjunction with the thermodynamic Bethe ansatz equations, can be utilizied to
calculate any physical quantity that pertains to the energy spectrum. There are constraints on the
quantum numbers in Takahashi's equations which imply particular selection rules
that determine the permitted combinations of elementary excitations and hence
the physical excitation spectrum \cite{DEGKKK00}.  Takahashi's equations can also be used as the starting point for the calculation of the scattering matrix
of the elementary excitations. For the half-filled Hubbard model in vanishing
magnetic field the $S$-matrix was calculated \cite{EK94,EK94+}.  The excitation spectrum at half filling is given by scattering states of
four elementary excitations: holon and antiholon with spin $0$ and charge $\pm
e$ and charge neutral spinons with spin up or down, respectively. This is
noteworthy, because away from half filling, or at finite magnetic field, the
number of elementary excitations is infinite \cite{DEGKKK00}. Furthermore the four particles can only be excited in $SO\left(4\right)$
multiplets \cite{EK94,EK94+,EFGKK05}.

In $1986$ Shastry started a novel way for studying the Hubbard model by
embedding it into the framework of the quantum inverse scattering method. By use of 
a Jordan-Wigner transformation he mapped the Hubbard model to a spin model and
then showed that the resulting spin Hamiltonian commutes with the
row-to-row transfer matrix of a related covering vertex model
\cite{Sha86+}. In this way, Shastry discovered the $R$-matrix of the Hubbard
model, thus placing it into the general context of integrable models
\cite{Sha86}, however with non-difference type spectral parameters. Later, it
was exhaustively shown that Shastry's $R$-matrix satisfies the Yang-Baxter
equation \cite{SW95}. The $S$-matrix was calculated by Andrei \cite{Andrei95}. An algebraic Bethe ansatz for the Hubbard model was
developed and expressions for the eigenvalues of the row-to-row transfer
matrix of the two-dimensional statistical covering model were calculated
\cite{RM97,YD97,MR98}. This was of significant importance for the
column-to-column transfer matrix (quantum transfer matrix, QTM) approach to
the thermodynamics of the Hubbard model \cite{JKS98}. This method grants an extremely simplified description of the thermodynamics in terms of the
solution of a finite set of non-linear integral equations, rather than the
infinite set, which was derived by Takahashi in $1972$ \cite{Tak72}. Thermodynamic quantities can be obtained numerically with
very high precision within
the \QTM\ approach.  Furthermore, the method can be used for the
calculation of correlation lengths \cite{Tsune91,USK03,EFGKK05}. The equivalence of QTM and TBA approach was shown in \cite{CCMT15}. 

The goal of this paper is the construction and investigation of a Hubbard
model with integrable impurity. The motivation for this research is
two-fold. First, the procedure of Andrei and Johannesson \cite{AJ84} by use of
commuting transfer matrices with inhomogeneities yields a clear construction
principle of interesting quantum chains with impurities. To the best of our
knowledge, it has been applied to the Hubbard model by Zvyagin and
Schlottmann \cite{ZS97}, however with a special impurity coupling parameter and a TBA treatment. The analysis of the general case and the derivation of a useful framework for the calculation of
the thermodynamical properties of the impurity are the main result of this
paper. Second, the Hubbard model with impurity allows for a non-trivial
continuum limit with vanishing bulk interaction, but non-zero impurity
interaction. In fact, the (integrable) Anderson impurity model can be
understood as a derivative of the (integrable) Hubbard model with
impurity. The detailed study of the continuum limit of this Hubbard model has
to be presented in a separate publication though.

The paper is organized as follows. In Sect.~II we review the Hubbard
Hamiltonian, Shastry's $R$-matrix and introduce the family of commuting
transfer matrices with inhomogeneity and derive the Bethe ansatz equations for
the Hamiltonian with impurity. Sect.~III is devoted to the thermodynamical
calculations on the basis of the \QTM. In order to make this paper
self-contained we review some of the necessary elements of the treatment by
finitely many non-linear integral equations. This is necessary for sketching
the computation of the leading eigenstate's eigenvalue function for general
spectral parameter which has not been done so far. A summary of this work is
given in Sect.~IV. The appendix contains explicit expressions
  for the Hamiltonian, an analytic low-temperature treatment of the impurity
  in the half-filled case, and an alternative treatment to Sect.~III by use of
  TBA.

\section{\label{sec:level1}Bethe ansatz equations of the Hubbard model with
  integrable impurity}

First, we review the essential characteristics of the bulk Hamiltonian
of the Hubbard model and its exactly solvable classical analogue in
two dimensions \cite{EFGKK05}. The Hubbard model describes 
lattice electrons on $L$ sites with hopping,
on-site Coulomb repulsion $U$ and external fields $\mu$ and $B$:
\begin{align}
H_{\text{Hubbard}} & =-\sum_{j=1}^{L}\left(\sum_{a=\uparrow,\downarrow}\left(c_{j+1,a}^{\dagger}c_{j,a}+c_{j,a}^{\dagger}c_{j+1,a}\right)-U\left(n_{j,\uparrow}-\frac{1}{2}\right)\left(n_{j,\downarrow}-\frac{1}{2}\right)\right.\nonumber \\
 & \qquad\qquad\;\left.+\mu\left(n_{j,\uparrow}+n_{j,\downarrow}\right)+B\left(n_{j,\uparrow}-n_{j,\downarrow}\right)\vphantom{{\sum_{a}\frac{1}{2}}}\right).\label{eq:H Hubbard}
\end{align}
For our purpose the global Hilbert space can be viewed as a product of local spaces corresponding and indexed by site $j$ and an additional spin label
$a=\uparrow,\downarrow$.
The classical analogue in two dimensions is defined on a double-layer square
lattice, consisting of $\uparrow$- and $\downarrow$-sublattices.  On each sublattice a six-vertex model of free fermion
type is defined with $R$-matrices denoted by $r_{\uparrow}$ and
$r_{\downarrow}$ which are coupled to a non-difference type matrix $R(\lambda,\mu)$
\cite{Sha88}
\begin{align}
R(\lambda,\mu) & =\cos(\lambda+\mu)\text{ch}\left(h(\lambda)-h(\mu)\right)r(\lambda-\mu)\nonumber \\
 & \quad+\cos(\lambda-\mu)\text{sh}\left(h(\lambda)-h(\mu)\right)r(\lambda+\mu)\sigma_{1,\uparrow}^{z}\sigma_{1,\downarrow}^{z},\label{eq:R Hubbard}\\
r_{a}(\lambda) & =\frac{\cos\lambda+\sin\lambda}{2}+\frac{\cos\lambda-\sin\lambda}{2}\sigma_{1,a}^{z}\sigma_{2,a}^{z}+\sigma_{1,a}^{+}\sigma_{2,a}^{-}+\sigma_{1,a}^{-}\sigma_{2,a}^{+},\nonumber \\
r(\lambda) & =r_{\uparrow}(\lambda)r_{\downarrow}(\lambda),\nonumber \\
\text{sh}\left(2h(\lambda)\right) & :=-\frac{U}{4}\sin\left(2\lambda\right),\nonumber 
\end{align}
where in contrast to \cite{Sha88}, the sign of $U$ has been changed so that the logarithmic derivative of the row-to-row transfer matrix yields the repulsive Hubbard model. This $R$-matrix satisfies the Yang-Baxter equation as shown in
\cite{SW95}. The $R$-matrix also satisfies a unitarity condition: A
product of two $R$-matrices reduces to the identity matrix times a
function of the spectral parameters. This function may be dropped by arranging
for a suitable normalization factor. We do not do this, but have to remember
this factor when mapping the Hamiltonian at finite temperature to a classical model. We define state vectors by
\[
\left|1\right\rangle =\left|+,-\right\rangle ,\qquad\left|2\right\rangle =\left|+,+\right\rangle ,\qquad\left|3\right\rangle =\left|-,-\right\rangle ,\qquad\left|4\right\rangle =\left|-,+\right\rangle ,
\]
where $\left|\sigma_\uparrow,\sigma_\downarrow\right\rangle$ corresponds to a
site occupied by a $\uparrow$ ($\downarrow$) particle / hole if
$\sigma_\uparrow$ ($\sigma_\downarrow$) is $-$ / $+$.

The row-to-row transfer matrix (\ref{eq:t_RTR}) is defined by a product of $L$ many
$R$-matrices with $\lambda$ ($0$) for the first (second) argument
corresponding to the auxiliary (quantum) space
\begin{equation}
t(\lambda)=\text{tr}_{\text{aux}}\bigotimes^{L}R(\lambda,0).\label{eq:t_RTR}
\end{equation}
In dependence on $\lambda$, this is a family of commuting transfer matrices which
reduces to a shift operator at $\lambda=0$ and its logarithmic derivative is
identical to the Hubbard Hamiltonian.

The Bethe ansatz eigenstates for the row-to-row transfer matrix and the
Hubbard Hamiltonian (\ref{eq:H Hubbard}) for $K$ electrons and $M$ down spins
are characterized by two sets of quantum numbers $\left\{ k_{j}\right\}
_{j=1}^{K}$ and $\left\{ \varLambda_{l}\right\} _{l=1}^{M}$, $2M\leq K\leq L$
which in general may be complex. They are known
as charge and spin rapidities, respectively. They satisfy the Lieb-Wu
equations \cite{LW68}
\begin{align*}
\e^{\i k_{j}L} & =\prod_{l=1}^{M}\frac{\varLambda_{l}-\sin k_{j}-i\frac{U}{4}}{\varLambda_{l}-\sin k_{j}+i\frac{U}{4}},\\
\prod_{j=1}^{K}\frac{\varLambda_{l}-\sin k_{j}-i\frac{U}{4}}{\varLambda_{l}-\sin k_{j}+i\frac{U}{4}} & =-\prod_{m=1}^{M}\frac{\varLambda_{l}-\varLambda_{m}-i\frac{U}{2}}{\varLambda_{l}-\varLambda_{m}+i\frac{U}{2}}.
\end{align*}
$\phantom{}$

$\phantom{}$

Next we study a row-to-row transfer matrix (\ref{eq:t_RTR}) defined on $L+1$
sites similar to above with a host of $L$ factors $R(\lambda,0)$ and an
impurity site with factor $R(\lambda,\nu)$
\begin{equation}
t(\lambda)=\text{tr}_{\text{aux}}\left[\bigotimes^{L}R\left(\lambda,0\right)\otimes
R\left(\lambda,\nu\right)\right],\label{eq:QTM_RTR}
\end{equation}
where $\nu$ is associated with site $L+1$ and may take arbitrary real or
complex values. Note that the model in \cite{ZS97} is obtained by choice of small $\nu$.  By construction, also this $t(\lambda)$ is a family of
commuting transfer matrices. The logarithmic derivative at $\lambda=0$ is a
sum of local terms, one site and two site operators for the bulk (\ref{eq:H Hubbard}) and a three
site impurity interaction.

At this point, we like to comment on the concrete form of the impurity
interaction with the host. It is relatively straightforward to derive the
following expression
\begin{equation}
h_{l,i,r}=\left[\partial_\nu R_{l,i}\left(\nu,0\right)\right] R_{i,l}\left(0,\nu\right) +
R_{l,i}\left(\nu,0\right) h_{l,r} R_{i,l}\left(0,\nu\right),\label{eq:h_imp}
\end{equation}
where we indexed the impurity site and the neighbouring ones by $l, i, r$
(left, impurity, right) instead of the above $L, L+1, 1$ in the construction
of the commuting family of transfer matrices and $h_{l,r}$ denotes the
standard local Hubbard interaction of the sites $l, r$. 

In analogy we consider the local Boltzmann weight associated with a vertex configuration $R_{\beta\delta}^{\alpha\gamma}\left(\lambda,\mu\right)$ and introduce $\bar{R}_{\beta\delta}^{\alpha\gamma}\left(\lambda,\mu\right)=R_{\delta\alpha}^{\gamma\beta}\left(\mu,\lambda\right)$ by clockwise $90^{\circ}$ rotations of $R\left(\lambda,\mu\right)$. Introduction of an auxiliary transfer matrix 
\[
\bar{t}\left(\lambda\right)=\text{tr}_{\text{aux}}\left[\bigotimes^{L}\bar{R}\left(\lambda,0\right)\otimes\bar{R}\left(\lambda,\nu\right)\right]
\]
yields
\begin{equation*}
\bar{h}_{l,i,r}=\left[\partial_\nu \bar{R}_{r,i}\left(\nu,0\right)\right] \bar{R}_{i,r}\left(0,\nu\right) +
\bar{R}_{r,i}\left(\nu,0\right) h_{r,l} \bar{R}_{i,r}\left(0,\nu\right).
\end{equation*}
The combination of the two transfer matrices $t\left(\lambda\right)$ and $\bar{t}\left(\lambda\right)$ provides a hermitian Hamiltonian. There are various ways of rewriting the impurity Hamiltonian in more explicit
terms. We may do so by use of creation and annihilation operators of electrons
on lattice sites. The expressions are given in appendix A. The Hamiltonian of our model is then given by 
\begin{align}
H= & -\sum_{j=1}^{L-1}\left(\sum_{a=\uparrow,\downarrow}\left(c_{j+1,a}^{\dagger}c_{j,a}+c_{j,a}^{\dagger}c_{j+1,a}\right)-U\left(n_{j,\uparrow}-\frac{1}{2}\right)\left(n_{j,\downarrow}-\frac{1}{2}\right)\right)\nonumber \\
 & -\sum_{j=1}^{L+1}\left(\mu\left(n_{j,\uparrow}+n_{j,\downarrow}\right)+B\left(n_{j,\uparrow}-n_{j,\downarrow}\right)\right)+h_{\text{imp}}.\label{eq:Hamiltonian}
\end{align} 
Another choice is the use of the momentum
representation. Our main application (in a later publication) will be the study of a suitable continuum
limit leading to the Anderson impurity model. For this application the
momentum representation is much more useful. The computational details however
require a separate publication.

\begin{figure}
\centering{}\includegraphics[scale=0.2]{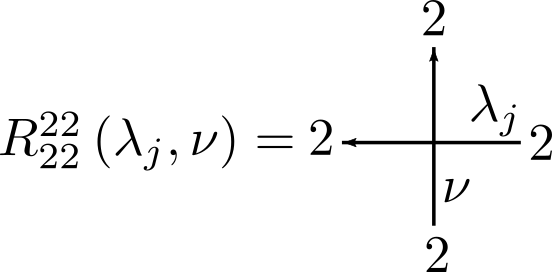}$\qquad$\includegraphics[scale=0.2]{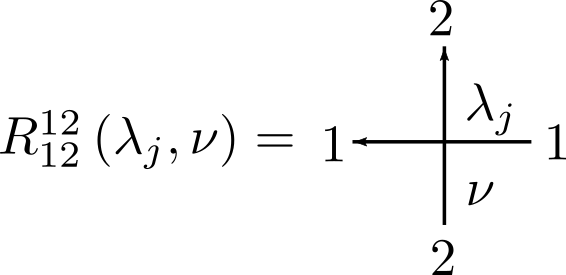}
\caption{Depiction
  of the two configurations of the impurity vertex that enter the eigenvalue
  equations. Note that ``$2$'' corresponds to the local vacuum and ``$1$''
  corresponds to the occupation with a single spin down electron.}
\label{impurity}
\end{figure}

The integrable impurity changes the Lieb-Wu equations by the impurity vertex shown in Fig.~\ref{impurity}.
which provides an additional phase factor
\begin{align*}
\frac{R_{22}^{22}\left(\lambda_{j},\nu\right)}{R_{12}^{12}\left(\lambda_{j},\nu\right)}
&
=\e^{2h(\nu)}\frac{\frac{z_{-}(\lambda_{j})}{z_{+}(\nu)}+1}{z_{-}(\nu)-z_{-}(\lambda_{j})}
=\e^{2h(\nu)}\frac{\e^{\i k_{j}}/{z_{+}(\nu)}+1}{z_{-}(\nu)-\e^{\i k_{j}}},
\end{align*}
where $\lambda_j$ and $k_j$ are different parameterizations of the charge
momenta: $z_{-}(\lambda_{j})=\e^{\i k_{j}}$.
Here we have used the functions $z_{\pm}(\lambda)$ and
$h(\lambda)$ which are defined by
\begin{equation}
z_{\pm}(\lambda):= \e^{2h(\lambda)\pm2x},\quad
\e^{2x}=\tan\lambda,\quad 
2h(\lambda)=
-\text{arsinh}\frac{U}{4\text{ch}\left(2x\right)}.\label{eq:Parametrisierung}
\end{equation}
The Bethe ansatz equations for the
row-to-row transfer matrix of the Hubbard model with impurity are thus
\begin{align}
\e^{\i k_{j}L}\e^{2h(\nu)}\frac{\e^{\i k_{j}}/{z_{+}(\nu)}+1}{z_{-}(\nu)-\e^{\i k_{j}}} & =\prod_{l=1}^{M}\frac{\varLambda_{l}-\sin k_{j}-i\frac{U}{4}}{\varLambda_{l}-\sin k_{j}+i\frac{U}{4}},\nonumber \\
\prod_{j=1}^{K}\frac{\varLambda_{l}-\sin k_{j}-i\frac{U}{4}}{\varLambda_{l}-\sin k_{j}+i\frac{U}{4}} & =-\prod_{m=1}^{M}\frac{\varLambda_{l}-\varLambda_{m}-i\frac{U}{2}}{\varLambda_{l}-\varLambda_{m}+i\frac{U}{2}}.\label{eq:TM BAE Hubbard + imp}
\end{align}
Note that for $\nu\to 0$ these equations reduce to the standard Lieb-Wu
equations with $L\to L+1$ as the additional phase factor on the l.h.s. of the
first equation turns into $\e^{\i k_{j}}$. The eigenvalue of the Hamiltonian of our model is given by
\[
E=-2\sum_{j=1}^{K}\cos k_{j}+\frac{U}{4}\left(L-2K\right)-\mu K-B\left(K-2M\right).
\]
\section{\label{sec:level1}Diagonalization of the column-to-column transfer matrix of the Hubbard model}

In this section we will treat the thermodynamical properties by mapping the
quantum Hamiltonian in one spatial dimension at finite temperature to a
classical model in two dimensions. In the Hamiltonian limit (small $\lambda$),
the transfer matrix (\ref{eq:t_RTR}) takes the form of an exponential of the
Hamiltonian times a translation operator by one lattice site. A certain
adjoint version of (\ref{eq:t_RTR}) with rotated vertices enjoys a similar
Hamiltonian limit, however with inverse translation operator. Therefore the
product of these two transfer matrices for small spectral parameter yields an
exponential expression of the Hamiltonian. This is still the case for the
transfer matrix with impurity (\ref{eq:QTM_RTR}) and its adjoint.

\begin{figure}
\begin{centering}
\includegraphics[scale=0.3]{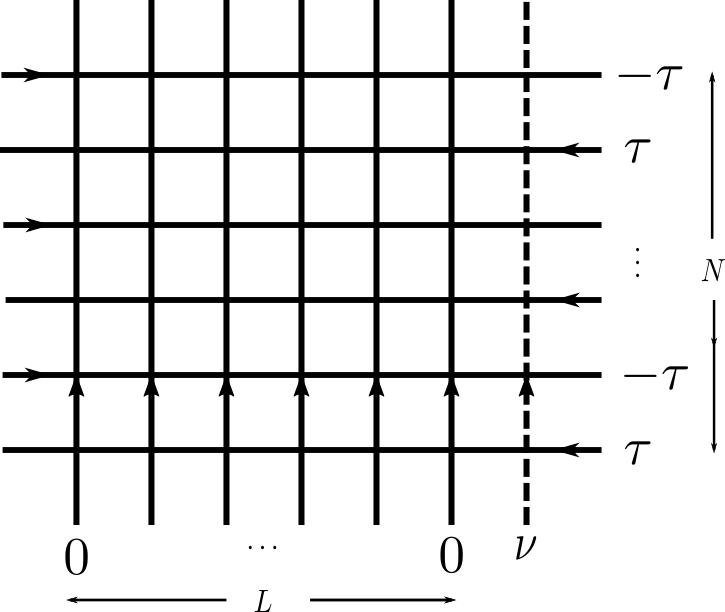}
\par\end{centering}
\caption{The quantum chain at finite temperature is mapped onto this two-dimensional
classical model. The square lattice has width $L+1$ equal to the
chain length, and height identical to the Trotter number $N$. The
alternating rows of the lattice correspond to two types of transfer matrices, where
$\tau=\frac{\beta}{N}$. The row-to-row transfer matrices commute. The column-to-column transfer matrices for the host
(black) and for the impurity (dashed line) also commute. The leading joint eigenstate and the
corresponding eigenvalues for the host and for the impurity yield the total
thermodynamical potential in the thermodynamic limit.}
\label{Trotter}
\end{figure}

The thermodynamical potential of the Hamiltonian with impurity is therefore
calculated from the partition function of the classical two-dimensional model
illustrated in Fig.~\ref{Trotter}.
In this section we use the QTM approach and the technique of
finitely many non-linear integral equations \cite{JKS98,EFGKK05}.
We want to calculate the impurity contribution to the thermodynamical potential.

Following \cite{JKS98} we introduce the column-to-column transfer matrix
\[
t^{\text{QTM}}\left(\lambda,\tau\right)=\text{tr}_{\text{aux}}\left(\bigotimes^{\frac{N}{2}}R\left(\lambda,-\tau\right)\otimes\tilde{R}\left(\lambda,\tau\right)\right),
\]
where $\tilde{R}$ is closely related to $R$.
The diagonalization of the column-to-column transfer matrix \cite{JKS98} is
algebraically very similar to that of the row-to-row case as the
column-to-column and the row-to-row transfer matrices have the same
intertwining operator. We use periodic or twisted boundary conditions in
Trotter direction, since we allow for an external magnetic field $B$ and a
chemical potential $\mu$.

A convenient vacuum is
\[
\left|\varOmega\right\rangle =\left|1,4,1,4,\ldots,1,4\right\rangle. 
\]
The vacuum expectation values are given by 
\begin{equation}
\left\langle \varOmega\right|t_{j,j}^{\text{QTM}}\left|\Omega\right\rangle =A_{j}\e^{\beta\mu_{j}},\qquad j=1,\ldots,4\label{eq:Ai}
\end{equation}
with 
\[
\mu_{1}=\mu+B,\qquad\mu_{2}=2\mu,\qquad\mu_{3}=0,\qquad\mu_{4}=\mu-B.
\]
We parameterize $\lambda,\tau$ in terms of $x, w$
\begin{equation}
\e^{2x}=\tan\lambda,\qquad \e^{2w}=\tan\tau\label{eq:Parametrisierung QTM}
\end{equation}
and use the functions that appeared already in (\ref{eq:Parametrisierung}).
The vacuum expectation values can be written as \cite{JKS98}
\begin{align}
\frac{A_{1}}{A_{2}} & =\left(\frac{\left(1-z_{-}\left(w\right)z_{+}(x)\right)\left(1-z_{+}\left(w\right)z_{+}(x)\right)}{\left(1+z_{-}\left(w\right)z_{+}(x)\right)\left(1+z_{+}\left(w\right)z_{+}(x)\right)}\right)^{\frac{N}{2}},\nonumber \\
\frac{A_{4}}{A_{2}} & =\left(\frac{\left(1+\frac{z_{-}\left(w\right)}{z_{-}(x)}\right)\left(1+\frac{z_{+}\left(w\right)}{z_{-}(x)}\right)}{\left(1-\frac{z_{-}\left(w\right)}{z_{-}(x)}\right)\left(1-\frac{z_{+}\left(w\right)}{z_{-}(x)}\right)}\right)^{\frac{N}{2}},\nonumber \\
A_{2} & =\left(\cos^{2}\lambda\cos^{2}\tau\cos^{2}\left(\lambda-\tau\right)\cos^{2}\left(\lambda+\tau\right)\e^{2h\left(w\right)}\left(\frac{1}{z_{-}\left(w\right)}-\frac{1}{z_{-}(x)}\right)\right.\nonumber \\
 & \quad\;\;\,\left.\cdot\left(z_{+}(x)+\frac{1}{z_{-}\left(w\right)}\right)\right)^{\frac{N}{2}},\nonumber \\
A_{3} & =A_{2}.\label{eq:A2}
\end{align}
The eigenvalues of the column-to-column transfer matrix are given by \cite{JKS98}
\begin{align}
\frac{\varLambda^{\text{QTM}}(\lambda)}{A_{2}} & =\e^{\beta\left(\mu+B\right)}\frac{A_{1}}{A_{2}}\prod_{j=1}^{m}\e^{2x}\frac{1+z_{j}z_{-}(x)}{1-z_{j}z_{+}(x)}\nonumber \\
 & \quad+\e^{2\beta\mu}\prod_{j=1}^{m}\left(-\e^{2x}\frac{1+z_{j}z_{-}(x)}{1-z_{j}z_{+}(x)}\right)\prod_{\alpha=1}^{l}\left(-\frac{z_{-}(x)-\frac{1}{z_{-}(x)}-2iw_{\alpha}+\frac{3U}{2}}{z_{-}(x)-\frac{1}{z_{-}(x)}-2iw_{\alpha}+\frac{U}{2}}\right)\nonumber \\
 & \quad+\prod_{j=1}^{m}\left(-\e^{-2x}\frac{1+\frac{z_{+}(x)}{z_{j}}}{1-\frac{z_{-}(x)}{z_{j}}}\right)\prod_{\alpha=1}^{l}\left(-\frac{z_{-}(x)-\frac{1}{z_{-}(x)}-2iw_{\alpha}-\frac{U}{2}}{z_{-}(x)-\frac{1}{z_{-}(x)}-2iw_{\alpha}+\frac{U}{2}}\right)\nonumber \\
 & \quad+\e^{\beta\left(\mu-B\right)}\frac{A_{4}}{A_{2}}\prod_{j=1}^{m}\e^{-2x}\frac{1+\frac{z_{+}(x)}{z_{j}}}{1-\frac{z_{-}(x)}{z_{j}}}\label{eq:EW Hubbard komp}
\end{align}
with rapidities $z_1,...,z_m$ and $w_1,...,w_l$. For the leading eigenvalue we
have to choose  $m = N$ and $l = N / 2$.

The parameters $\left\{ z_{j}\right\} _{j=1}^{m}$ and $\left\{
w_{\alpha}\right\} _{\alpha=1}^{l}$ are determined from the Bethe ansatz equations
\begin{align}
\e^{\beta\left(\mu-B\right)}\left(\frac{\left(1+\frac{z_{-}\left(w\right)}{z_{j}}\right)\left(1+\frac{z_{+}\left(w\right)}{z_{j}}\right)}{\left(1-\frac{z_{-}\left(w\right)}{z_{j}}\right)\left(1-\frac{z_{+}\left(w\right)}{z_{j}}\right)}\right)^{\frac{N}{2}} & =\left(-1\right)^{1+m}\prod_{\alpha=1}^{l}\left(-\frac{z_{j}-\frac{1}{z_{j}}-2iw_{\alpha}-\frac{U}{2}}{z_{j}-\frac{1}{z_{j}}-2iw_{\alpha}+\frac{U}{2}}\right),\nonumber \\
\e^{2\beta\mu}\prod_{\alpha=1}^{l}\frac{z_{j}-\frac{1}{z_{j}}-2iw_{\alpha}+\frac{U}{2}}{z_{j}-\frac{1}{z_{j}}-2iw_{\alpha}-\frac{U}{2}} & =-\prod_{\beta=1}^{l}\frac{2i\left(w_{\alpha}-w_{\beta}\right)-U}{2i\left(w_{\alpha}-w_{\beta}\right)+U}.\label{eq:Hubbard BAE 2 Form 1}
\end{align}
In the limit $U\rightarrow0$ we find the free-fermion partition function. We may
use the alternative vacuum $\left|\varOmega^{\prime}\right\rangle
=\left|2,3,2,3,\ldots,2,3\right\rangle $, for which we find another formula for
$\varLambda^{\text{QTM}}(\lambda)$.  This is the same as equation (\ref{eq:EW
  Hubbard komp}) after changing the sign of $U$ and swapping
$B\longleftrightarrow\mu$ which can be understood as a partial particle-hole
transformation. The solutions of the Bethe ansatz equations of the
column-to-column transfer matrix (\ref{eq:Hubbard BAE 2 Form 1}) for the
leading eigenvalue $\varLambda^{\text{QTM}}(\lambda)$ have a characteristic
temperature dependence. For high temperatures $T$ all $z_{j}$ satisfy
$\text{Re}z_{j}=0$ and $\left|z_{j}\right|>1$.  Lowering the temperature $T$
yields a decrease of the $\left|z_{j}\right|$ and they converge to the origin. For low
temperatures $T$ a certain number of the $z_{j}$'s satisfy
$\left|z_{j}\right|<1$.  The $w_{\alpha}$ parameters behave alike
  on the real axis.


In order to uniformize the Bethe ansatz equations (\ref{eq:Hubbard BAE 2 Form
  1}) we use the function $s\left(z\right)$ (whose inverse is a double valued function
with branch cut from $-1$ to $1$)
\begin{align}
s(z)=\frac{1}{2i}\left(z-\frac{1}{z}\right),\qquad
s_{j} =\frac{1}{2i}\left(z_{j}-\frac{1}{z_{j}}\right),\label{eq:s Variable}
\end{align}
and express the $z_{j}$'s in terms of $s_{j}$ parameters. Note that $z$ values
satisfying $\text{Re}z=0$ are mapped onto the same area of the real axis with
$\left|s\right|>1$, regardless wether $\left|z\right|>1$ or $\left|z\right|<1$
holds. The above described motion of $z_j$ parameters upon lowering the
temperature leads to a motion of the $s_{j}$ parameters from the first branch
to the second branch. At high temperatures $T$, all parameters $s_{j}$ lie on
the first sheet of the complex plane. At low temperatures $T$, the parameters
$s_{j}$ lie on the first and on the second sheet.

\subsection{\label{sec:level2}Associated auxiliary problem of difference type}

We want to reformulate the Bethe ansatz equations (\ref{eq:Hubbard BAE 2 Form
  1}) of the column-to-column transfer matrix in the limit
$N\rightarrow\infty$ as a system of non-linear integral equations. First, the
Bethe ansatz equations (\ref{eq:Hubbard BAE 2 Form 1}) can be written in
difference form in the rapidities $\left\{ s_{j}\right\} _{j=1}^{m}$ and
$\left\{ w_{\alpha}\right\} _{\alpha=1}^{l}$
\begin{align}
\e^{-\beta\left(\mu-B\right)}\Phi\left(s_{j}\right) & =-\frac{q_{2}\left(s_{j}-i\frac{U}{4}\right)}{q_{2}\left(s_{j}+i\frac{U}{4}\right)},\nonumber \\
\e^{-2\beta\mu}\frac{q_{2}\left(w_{\alpha}+i\frac{U}{2}\right)}{q_{2}\left(w_{\alpha}-i\frac{U}{2}\right)} & =-\frac{q_{1}\left(w_{\alpha}+i\frac{U}{4}\right)}{q_{1}\left(w_{\alpha}-i\frac{U}{4}\right)},\label{eq:Hubbard BAE 2 Form 2}
\end{align}
where we have defined 
\begin{align}
q_{1}(s) & :=\prod_{j=1}^{m}\left(s-s_{j}\right),\qquad q_{2}:=\prod_{\alpha=1}^{l}\left(s-w_{\alpha}\right),\label{eq:q Hubbard}\\
\Phi(s) & :=\left(\frac{\left(1-\frac{z_{-}\left(w\right)}{z(s)}\right)\left(1-\frac{z_{+}\left(w\right)}{z(s)}\right)}{\left(1+\frac{z_{-}\left(w\right)}{z(s)}\right)\left(1+\frac{z_{+}\left(w\right)}{z(s)}\right)}\right)^{\frac{N}{2}},\label{eq:Phi Hubbard}\\
z(s) & :=is\left(1+\sqrt{1-\frac{1}{s^{2}}}\right).\label{eq:z Hubbard}
\end{align}
Note that the functions $\Phi(s)$ and $z(s)$
have two branches: The requirement $z(s)\simeq2is$ for
large values of $s$ defines the standard first branch of $z(s)$.
The branch cut of $z(s)$ for values of $z$ on the unit
circle is along $\left[-1,1\right]$. Thus the first branch of the
function $z(s)$ maps the complex plane without $\left[-1,1\right]$
to the outer area of the complex plane of the unit circle. Vice versa
the second branch of $z(s)$ maps the complex plane without
$\left[-1,1\right]$ to the inner area of the unit circle. On the branch cut we have 
\[
z\left(s\pm i0\right)=is\mp\sqrt{1-s^{2}},\qquad s\in\left[-1,1\right].
\]
The two branches of the function $\Phi(s)$ defined in
(\ref{eq:Phi Hubbard}) are denoted by $\Phi^{\pm}(s)$.
The function $\Phi^{+}(s)$ has a zero (pole) of order
$\frac{N}{2}$ at the point $s_{0}$ ($-s_{0}$). $\Phi^{-}(s)$
has a zero (pole) of order $\frac{N}{2}$ at the point $-s_{0}+i\frac{U}{2}$
($s_{0}-i\frac{U}{2}$). The point $s_{0}$ is defined by
$
z\left(s_{0}\right):=z_{-}\left(w\right).
$
The general expression for the leading eigenvalue $\varLambda^{\text{QTM}}(\lambda)$
(\ref{eq:EW Hubbard komp}) is quite complicated, but simplifies considerably
by use of the relations
\begin{align*}
z_{+}(x)-\frac{1}{z_{+}(x)}+z_{-}(x)-\frac{1}{z_{-}(x)} & =-U,\\
\left(1+\frac{z_{+}(x)}{z_{j}}\right)\left(1-z_{j}z_{+}(x)\right) & =2iz_{+}(x)\left(s-s_{j}-i\frac{U}{2}\right),\\
\left(1+z_{j}z_{-}(x)\right)\left(1-\frac{z_{-}(x)}{z_{j}}\right) & =2iz_{-}(x)\left(s_{j}-s\right)
\end{align*}
and by use of the functions 
\begin{align}
\lambda_{1}(s) & =\e^{\beta\left(\mu+B\right)}\frac{\Phi\left(s-i\frac{U}{4}\right)}{q_{1}\left(s-i\frac{U}{4}\right)},\qquad\lambda_{2}(s)=\e^{2\beta\mu}\frac{q_{2}\left(s-i\frac{U}{2}\right)}{q_{1}\left(s-i\frac{U}{4}\right)q_{2}(s)},\nonumber \\
\lambda_{3}(s) & =\frac{q_{2}\left(s+i\frac{U}{2}\right)}{q_{1}\left(s+i\frac{U}{4}\right)q_{2}(s)},\qquad\lambda_{4}(s)=\frac{\e^{\beta\left(\mu-B\right)}}{\Phi\left(s+i\frac{U}{4}\right)q_{1}\left(s+i\frac{U}{4}\right)},\nonumber \\
\Lambda^{\text{aux}}(s) & =\lambda_{1}(s)+\lambda_{2}(s)+\lambda_{3}(s)+\lambda_{4}(s).\label{eq:Lambda aux}
\end{align}
The r.h.s. of (\ref{eq:EW Hubbard komp}) can be written as a common factor
times the auxiliary function $\Lambda^{\text{aux}}$ for even $m$ and $l$
\begin{equation}
\frac{\varLambda^{\text{QTM}}(\lambda)}{A_{2}}=\left(\frac{\e^{-2h(x)}}{2i}\right)^{m}\Lambda^{\text{aux}}\left(s-i\frac{U}{4}\right)\prod_{j=1}^{m}\left(\left(1+z_{j}z_{-}(x)\right)\left(1+\frac{z_{+}(x)}{z_{j}}\right)\right).\label{eq:Hubbard Eigenwert Zwischenergebnis}
\end{equation}
Note that on the right-hand side $x$ and $s=s\left(z_{-}\left(x\right)\right)$ depend on $\lambda$
via (\ref{eq:Parametrisierung QTM}) and (\ref{eq:s Variable}).

The requirement of analyticity of $\Lambda^{\text{aux}}(s)$
yields the equations (\ref{eq:Hubbard BAE 2 Form 2}), which are the
Bethe ansatz equations of the (leading) eigenvalue $\varLambda^{\text{QTM}}(\lambda)$.
For the leading eigenvalue note that while $\varLambda^{\text{QTM}}(\lambda)$
is analytic everywhere, $\Lambda^{\text{aux}}(s)$ is analytic
on the first branch, but may have singularities on the other three
branches since there are two branch cuts at $\left[-1,1\right]\pm\i U/4$.

We remark that the functions $\left(\lambda_{1}+\lambda_{2}\right)(s)$,
$\left(\lambda_{3}+\lambda_{4}\right)(s)$ and $\Lambda^{\text{aux}}(s)$
have zero winding number around their branch cuts, because
the number of poles on the first branch is $N$ and the asymptotics
of the functions is ${1}/{s^{N}}$.

\subsection{\label{sec:level2}Non-linear integral equations}

We consider the integral equations equivalent to the nested Bethe ansatz
equations for the leading eigenvalue of the column-to-column transfer matrix
for $U>0$. We use a set of auxiliary functions satisfying a set of closed
non-linear integral equations. The following definitions are very useful:
\begin{align}
l_{j}(s) & :=\e^{2\beta B}\lambda_{j}\left(s-i\frac{U}{4}\right)\Phi^{+}(s)\Phi^{-}(s),\qquad j=1,\ldots,4,\nonumber \\
\bar{l}_{j}(s) & :=\lambda_{j}\left(s+i\frac{U}{4}\right),\qquad j=1,\ldots,4,\nonumber \\
\mathfrak{b}(s) & :=\frac{\bar{l}_{1}+\bar{l}_{2}+\bar{l}_{3}+\bar{l}_{4}}{l_{1}+l_{2}+l_{3}+l_{4}}(s),\nonumber \\
\bar{\mathfrak{b}}(s) & :=\frac{1}{\mathfrak{b}}(s),\nonumber \\
\mathfrak{c}(s) & :=\frac{\left(l_{1}+l_{2}\right)\left(\bar{l}_{1}+\bar{l}_{2}+\bar{l}_{3}+\bar{l}_{4}\right)}{\left(l_{3}+l_{4}\right)\left(l_{1}+l_{2}+l_{3}+l_{4}+\bar{l}_{1}+\bar{l}_{2}+\bar{l}_{3}+\bar{l}_{4}\right)}(s),\nonumber \\
\bar{\mathfrak{c}}(s) & :=\frac{\left(\bar{l}_{3}+\bar{l}_{4}\right)\left(l_{1}+l_{2}+l_{3}+l_{4}\right)}{\left(\bar{l}_{1}+\bar{l}_{2}\right)\left(l_{1}+l_{2}+l_{3}+l_{4}+\bar{l}_{1}+\bar{l}_{2}+\bar{l}_{3}+\bar{l}_{4}\right)}(s)\nonumber \\
\mathfrak{B}(s) & :=1+\mathfrak{b}(s),\nonumber \\
\bar{\mathfrak{B}}(s) & :=1+\bar{\mathfrak{b}}(s),\nonumber \\
\mathfrak{C}(s) & :=1+\mathfrak{c}(s),\nonumber \\
\bar{\mathfrak{C}}(s) & :=1+\bar{\mathfrak{c}}(s).\label{eq:Auxiliary functions}
\end{align}
We note that any analytic function on the complex plane is settled
by its singularities and its asymptotic behaviour at infinity. All
of the above defined auxiliary functions $\mathfrak{b}(s)$,
$\mathfrak{c}(s)$ and $\bar{\mathfrak{c}}(s)$
show constant asymptotics for finite $N$. By investigating the function $\lambda_{1}(s)+\lambda_{2}(s)+\lambda_{3}(s)+\lambda_{4}(s)$
we find poles of order $\frac{N}{2}$ at $s_{0}-i\frac{U}{4}$ and
$i\frac{U}{4}-s_{0}$. We also have zeroes and branch cuts on the
lines $\text{Im}\,s=\pm\frac{U}{4}$. This yields the following expression 
\begin{align*}
\ln\left(l_{1}(s)+l_{2}(s)+l_{3}(s)+l_{4}(s)\right) & \equiv_{s}-\frac{N}{2}\ln\left(\left(s-s_{0}\right)\left(s+s_{0}-i\frac{U}{2}\right)\right)\\
 & \quad\;\,+\ln\left(\Phi^{+}(s)\Phi^{-}(s)\right)+L_{-}(s)+L_{+}\left(s-i\frac{U}{2}\right),
\end{align*}
where $\equiv_{s}$ indicates that the left and right hand sides have
the same singularities on the entire plane and the functions $L_{\pm}$ are
defined by Cauchy integrals
\begin{align}
L_{\pm}(s) & =\left(k\circ l_{\pm}\right)(s),\nonumber \\
k(s) & =\frac{1}{2\pi is},\quad
l_{\pm}(s) =\left(\lambda_{1}+\lambda_{2}+\lambda_{3}+\lambda_{4}\right)\left(s\pm i\frac{U}{4}\right),
\label{eq:k Kern}\\
\left(g\circ f\right)(s) & =\intop_{\mathcal{L}}\mathrm{d}tg\left(s-t\right)f\left(t\right).\label{eq:Integrationskontur}
\end{align}
The contour $\mathcal{L}$ in the convolution integrals surrounds the real axis
at infinitesimal distance above and below in anticlockwise manner.

Using furthermore the identity 
\[
\Phi^{+}(s)\Phi^{-}(s)=\left(\frac{\left(s-s_{0}\right)\left(s+s_{0}-i\frac{U}{2}\right)}{\left(s+s_{0}\right)\left(s-s_{0}+i\frac{U}{2}\right)}\right)^{\frac{N}{2}},
\]
and the singularities of the functions $\ln\mathfrak{B}$ and $\ln\bar{\mathfrak{c}}-\ln\bar{\mathfrak{C}}$
we get 
\begin{align}
\ln\left(l_{1}(s)+l_{2}(s)+l_{3}(s)+l_{4}(s)\right) & \equiv_{s}-\frac{N}{2}\ln\left(\left(s+s_{0}\right)\left(s-s_{0}+i\frac{U}{2}\right)\right)\nonumber \\
 & \quad\;\,-\left(k\circ\ln\mathfrak{B}\right)(s)\nonumber \\
 & \quad\;\,+\left(k\circ\left(\ln\bar{\mathfrak{C}}-\ln\bar{\mathfrak{c}}-\ln\mathfrak{B}\right)\right)\left(s-i\frac{U}{2}\right).\label{eq:NLIE f=0000FCr Lambda auxiliary-1}
\end{align}
The asymptotic behaviour at infinity is given by 
\begin{align*}
\ln\mathfrak{b}(s) & \overset{s\rightarrow\infty}{\longrightarrow}-2\beta B,\\
\ln\bar{\mathfrak{c}}(s) & \overset{s\rightarrow\infty}{\longrightarrow}-\beta\left(\mu-B\right)-\ln\left(1+\e^{2\beta B}\right).
\end{align*}
For later use we define the kernel functions 
\begin{align}
K_{1}(s) & =k\left(s-i\frac{U}{4}\right)-k\left(s+i\frac{U}{4}\right)
=\frac{U}{4\pi}\frac{1}{s^{2}+\left(\frac{U}{4}\right)^{2}},\nonumber \\
\hat{K}_{1}(s) & =K_{1}\left(s+i\frac{U}{4}\right),\nonumber \\
\bar{K}_{1}(s) & =K_{1}\left(s-i\frac{U}{4}\right),\nonumber \\
K_{2}(s) & =k\left(s-i\frac{U}{2}\right)-k\left(s+i\frac{U}{2}\right)
=\frac{U}{2\pi}\frac{1}{s^{2}+\left(\frac{U}{2}\right)^{2}}.\label{eq:Integrationskerne Hubbard}
\end{align}
Next we note for convolutions of the kernel $k$ (with pole at $0$) and some
analytic function $f$ for a contour surrounding ($\oblong$) the argument $s$ 
and for a contour not surrounding ($\circ$) it:
\[
\left(k\oblong f\right)(s)=\left(k\circ f\right)(s)+f(s).
\]
For the wide integration contour we use a
loop around the real axis consisting of the two horizontal lines
$\text{Im}s=\pm\alpha$ with $0<\alpha\leq\frac{U}{4}$ and for the narrow
contour we use $\mathcal{L}$.

We find the following
non-linear integral equations for the auxiliary functions $\mathfrak{b}(s)$,
$\mathfrak{c}(s)$ and $\bar{\mathfrak{c}}(s)$ in the Trotter limit $N\rightarrow\infty$
\begin{align}
\ln\mathfrak{b}(s) & =-2\beta B+\left(K_{2}\oblong\ln\mathfrak{B}\right)(s)+\left(\bar{K}_{1}\circ\left(\ln\bar{\mathfrak{c}}-\ln\bar{\mathfrak{C}}\right)\right)(s),\nonumber \\
\ln\mathfrak{c}(s) & =-\frac{\beta U}{2}+\beta\left(\mu+B\right)-2i\beta s\sqrt{1-\frac{1}{s^{2}}}-\left(\bar{K}_{1}\oblong\ln\bar{\mathfrak{B}}\right)(s)-\left(\bar{K}_{1}\circ\ln\bar{\mathfrak{C}}\right)(s),\nonumber \\
\ln\bar{\mathfrak{c}}(s) & =-\frac{\beta U}{2}-\beta\left(\mu+B\right)+2i\beta s\sqrt{1-\frac{1}{s^{2}}}+\left(\hat{K}_{1}\oblong\ln\mathfrak{B}\right)(s)+\left(\hat{K}_{1}\circ\ln\mathfrak{C}\right)(s).\label{eq:NLIE Hubbard Form 1}
\end{align}
We note that the function $\mathfrak{b}(s)$ will be calculated
on the lines $\text{Im}\,s=\pm\alpha$, especially for $\alpha=\frac{U}{4}$.
The functions $\mathfrak{c}(s)$ and $\bar{\mathfrak{c}}(s)$
will be calculated on the real axis infinitesimally above and below
the interval $\left[-1,1\right]$. Note furthermore that these functions
are analytic outside of $\left[-1,1\right]$. Therefore convolutions
with these functions $\mathfrak{c}(s)$ and $\bar{\mathfrak{c}}(s)$
can be reduced to contours surrounding $\left[-1,1\right]$. We have
to solve the set of non-linear integral equations (\ref{eq:NLIE Hubbard Form 1})
for the auxiliary functions $\mathfrak{b}(s)$, $\mathfrak{c}(s)$
and $\bar{\mathfrak{c}}(s)$ before calculating the free
energy. 

\subsection{\label{sec:level2}Integral expression for the leading eigenvalue}

Here we present the derivation of the leading eigenvalue
$\varLambda^{\text{QTM}}(\lambda)$ of the column-to-column transfer matrix in
terms of the auxiliary functions (\ref{eq:NLIE Hubbard Form 1}). This
eigenvalue is known for $\lambda=0$ \cite{JKS98}, but not for general argument $\lambda$. We use a contour
$\mathcal{L}_{0}$ encircling the $s_{j}$ anticlockwise. The rapidities $s_{j}$
are not located on the branch cut of
$\ln\left(\left(1+z(s)z_{-}(x)\right)\left(1+\frac{z_{+}(x)}{z(s)}\right)\right)$
from $-1$ to $1$, therefore $\mathcal{L}_{0}$ consists of two disconnected
parts. For zero external fields these contours are loops around
$\left]-\infty,-1\right]$ and $\left[1,\infty\right[$, respectively. In the
        general case with non-zero external fields they are appropriately
        deformed. For the general case we use Cauchy's integral and write
\begin{align}
&g(t):=(1+z(t)z_{-}(x))\left(1+\frac{z_{+}(x)}{z(t)}\right),\quad
f(t):=\left[\ln g(t)\right] \left[\ln\left(1+\frac{l_{4}}{l_{3}}(t)\right)\right]^{\prime},\label{eq:Def g(s)}\\
&2\pi i\sum_{i=1}^{m}\ln\left(\left(1+z_{j}z_{-}(x)\right)\left(1+\frac{z_{+}(x)}{z_{j}}\right)\right)
=\nonumber\\
&=2\pi i\sum_{i=1}^{m}\ln g(s_j)=\underbrace{\intop_{\mathcal{L}_{0}}\mathrm{d}t\left.f(t)\right|_{1\text{st branch}}}_{=\varSigma_{1}}+\underbrace{\intop_{\mathcal{L}_{0}}\mathrm{d}t\left.f(t)\right|_{2\text{nd branch}}}_{=\varSigma_{2}},\label{eq:NR 5}
\end{align}
where $\varSigma_{1}$ and $\varSigma_{2}$ will be calculated below. 

\subsection{\label{sec:level2}Integral expression in terms of auxiliary functions}

\begin{figure}
\begin{centering}
\includegraphics[scale=0.3]{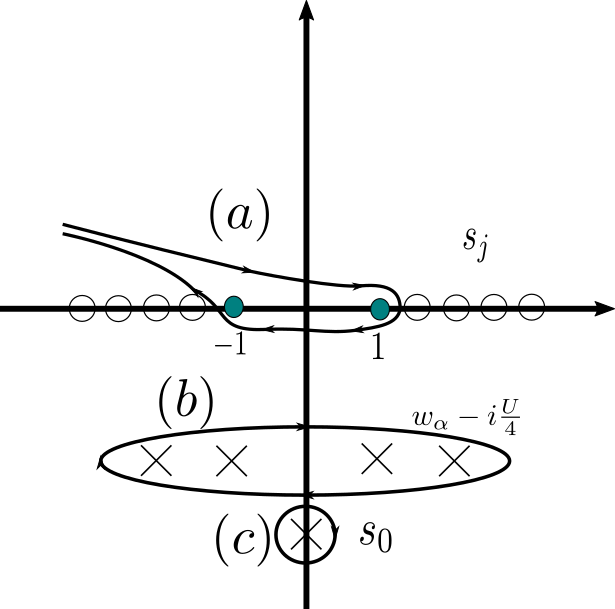}
\par\end{centering}
\caption{Zeroes and poles of the function $1+\frac{l_{4}}{l_{3}}(t)$:
Zeroes (poles) are illustrated by open circles (crosses) and are located
at $s_{j}$ ($w_{\alpha}-i\frac{U}{4}$). A pole is located at $s_{0}$
and has order $\frac{N}{2}$. Integration contour with three separate parts: $\left(a\right)$ is starting at $-\infty$, encircling
the interval $\left[-1,1\right]$ in clockwise manner and is going
back to $-\infty$. $\left(b\right)$ is a loop surrounding all $w_{\alpha}-i\frac{U}{4}$.
$\left(c\right)$ is a small circle arround $s_{0}$. }
\label{Contour1}
\end{figure}

\begin{figure}
\begin{centering}
\includegraphics[scale=0.3]{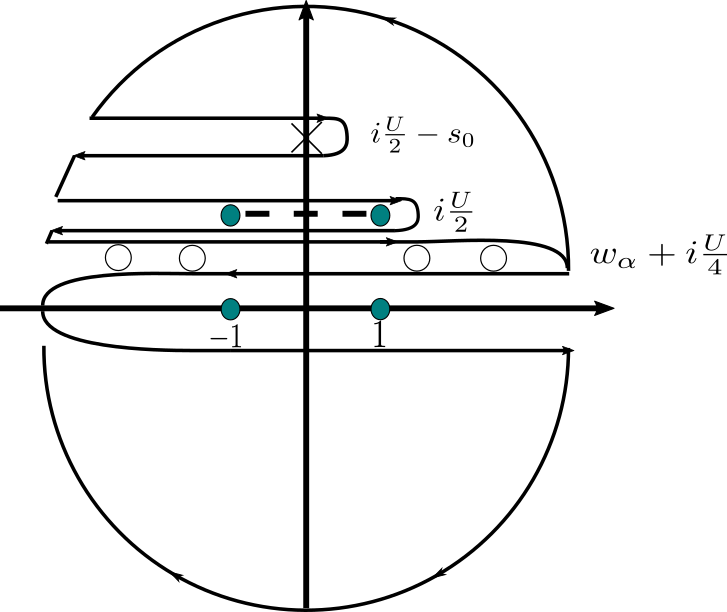}
\par\end{centering}
\caption{Depiction of zeroes and singularities of
  $\frac{l_{1}+l_{2}+l_{3}+l_{4}}{l_{3}+l_{4}}(t)$: zeroes are illustrated by
  open circles, branch cuts by dashed lines and the pole by a cross. We are
  interested in the integral (\ref{eq:Definition von Sigma}) along the contour $\mathcal{L}$
  which runs around the real axis from $\infty+\i\epsilon$ to
  $-\infty+\i\epsilon$, and then from $-\infty-\i\epsilon$ to
  $\infty-\i\epsilon$. We add a large semi-circle with radius $R$ to the lower
  half-plane. The integrand vanishes like $\mathcal{O}\left(\frac{\ln
    R}{R^{2}}\right)$ asymptotically. Therefore this path does not contribute
  to the integral (\ref{eq:Definition von Sigma}). Next we add a closed loop that does not
  encircle any singularities: this consists of a large semi-circle in the
  upper half-plane and indentations. Finally we drop the loop around the
  semi-disk in the lower half-plane and the semi-circle in the upper
  half-plane. This procedure keeps the integral unchanged, but transforms the
  contour into three indentations. The contours are around the
zeroes $w_{\alpha}+i\frac{U}{4}$, the branch cut $\left[-1,1\right]+i\frac{U}{2}$,
and the pole $i\frac{U}{2}-s_{0}$ of $\frac{l_{1}+l_{2}+l_{3}+l_{4}}{l_{3}+l_{4}}(t)$
in clockwise manner. The first and third contour integrals can be
calculated easily.
\label{Zeroes-and-Singularities}}
\end{figure}

The function $\frac{l_{4}}{l_{3}}(t)$ for $t\rightarrow\infty$
shows the asymptotic behaviour
$
\frac{l_{4}}{l_{3}}(t)\overset{t\rightarrow\infty}{\longrightarrow}\e^{\beta\left(\mu-B\right)}+\mathcal{O}\left(s^{-1}\right).
$
$z(t)$ is of order $\mathcal{O}(t)$.
Hence we add two large semi-circles with radius $R$ to the integration
contour $\mathcal{L}_{0}$ without changing the integral expression
of $\varSigma_{1}$. 
We deform the integration contour without changing the value of the
integral. As long as the contour does not run over singularities of $f(t)$ we
may do so according to Cauchy's theorem ($f(t)$ has a branch cut along the
interval $\left[-1,1\right]$ and poles that arise from zeroes and poles of
$1+\frac{l_{4}}{l_{3}}(t)$). This yields a contour with three separate parts, which are illustrated in Fig.~\ref{Contour1}.
Contour $\left(a\right)$ contains a path $\left(a_{1}\right)$ from
$-\infty$ to $-1$, a loop $\left(a_{2}\right)$ around the interval
$\left[-1,1\right]$ and a path $\left(a_{3}\right)$ from $-1$ back to
$-\infty$. The paths $\left(a_{1}\right)$ and $\left(a_{3}\right)$
are obviously inverse to each other.
The integrals on the parts $\left(b\right)$ and $\left(c\right)$
can be calculated.
Now we consider $\varSigma_{2}$ and deform the integration contour
$\mathcal{L}_{0}$ as above. Note that the integral of
$\left.f(t)\right|_{1\text{st branch}}$ on part $\left(a_{2}\right)$ is
equal to the integral of $\left.f(t)\right|_{2\text{nd branch}}$ along
$\left(a_{2}\right)$ in reversed sense. 
Now we join the results for $\Sigma_1$ and $\Sigma_2$ and obtain
\begin{equation}
\varSigma_{1}+\varSigma_{2}=2\pi i\left(\sum_{\alpha=1}^{l}\ln g\left(w_{\alpha}-i\frac{U}{4}\right)+\frac{N}{2}\ln g\left(s_{0}\right)\right)+\intop_{\mathcal{L}}\mathrm{d}t\left.f(t)\right|_{2\text{nd branch}}.\label{eq:NR 2}
\end{equation}
We want to rewrite this result in terms of integrals involving the auxiliary
functions (\ref{eq:NLIE Hubbard Form 1}). To this end we consider
\begin{align}
\varSigma & :=\intop_{\mathcal{L}}\mathrm{d}t\left[\ln g\left(t-i\frac{U}{2}\right)\right]^{\prime}\ln\mathfrak{C}(t)+\intop_{\mathcal{L}}\mathrm{d}t\left[\ln g(t)\right]^{\prime}\ln\frac{1+\mathfrak{\mathfrak{c}+\bar{\mathfrak{c}}}}{\bar{\mathfrak{c}}}(t).\label{eq:Definition von Sigma}
\end{align}
First, we integrate by parts and use that $\ln g\left(t-i\frac{U}{2}\right)$
and $\ln g(t)$ show no jump after surrounding the real
axis. Therefore the surface terms vanish.
Next we use the factorization
\begin{align*}
\mathfrak{C}(t) & =\left(\frac{\sum_{j=1}^{4}l_{j}}{l_{3}+l_{4}}\cdot\frac{l_{3}+l_{4}+\sum_{j=1}^{4}\bar{l}_{j}}{\sum_{j=1}^{4}\left(l_{j}+\bar{l}_{j}\right)}\right)(t),\\
\frac{1+\mathfrak{\mathfrak{c}+\bar{\mathfrak{c}}}}{\bar{\mathfrak{c}}}(t) & =\frac{\sum_{j=1}^{4}\bar{l}_{j}}{\bar{l}_{3}+\bar{l}_{4}}(t)\cdot\underbrace{\frac{l_{3}+l_{4}+\bar{l}_{1}+\bar{l}_{2}}{l_{3}+l_{4}}(t)}_{=\left.1+\frac{l_{4}}{l_{3}}(t)\right|_{2\text{nd branch}}}
\end{align*}
and the fact that the second fraction of $\mathfrak{C}(t)$
is analytic along the real axis. According to Cauchy's theorem
it vanishes. Furthermore we deform the integration contour $\mathcal{L}$. This yields a contour
with three separate parts, which are illustrated in Fig.~\ref{Zeroes-and-Singularities}. 

The second contour is equal to the contour $\mathcal{L}+i\frac{U}{2}$
in clockwise manner. We rewrite this contour integral by the use of
a shift of the integration variable from $t$ to $t+i\frac{U}{2}$.
Then we have to exchange $l_{j}(t)$ functions by $\bar{l}_{j}(t)$
functions. Then some integral contributions cancel each other
\begin{align}
\varSigma & =2\pi i\left(\sum_{\alpha=1}^{l}\ln g\left(w_{\alpha}-i\frac{U}{4}\right)-\frac{N}{2}\ln g\left(-s_{0}\right)\right)-\intop_{\mathcal{L}}\mathrm{d}t\ln g(t)\left[\left.\ln\left(1+\frac{l_{4}}{l_{3}}(t)\right)\right|_{2\text{nd branch}}\right]^{\prime}.\label{eq:NR 3}
\end{align}
Comparing (\ref{eq:NR 2}) with (\ref{eq:NR 3}) and using 
$
\left.z(t)\right|_{2\text{nd branch}}=-\frac{1}{\left.z(t)\right|_{1\text{st branch}}}
$
yields 
\begin{align}
\varSigma_{1}+\varSigma_{2} & =\varSigma+2\pi i\left(\frac{N}{2}\left(\ln g\left(s_{0}\right)+\ln g\left(-s_{0}\right)\right)\right)\nonumber \\
 & \quad+2\pi i \left[\left(k\circ\ln\left(1+\frac{l_{4}}{l_{3}}\right)\right)(s)+\left(k\circ\ln\left(1+\frac{l_{4}}{l_{3}}\right)\right) \left( s-i\frac{U}{2}\right) \right].\label{eq:NR 4}
\end{align}
Using 
\begin{align*}
\frac{l_{3}+l_{4}+\bar{l}_{1}+\bar{l}_{2}}{l_{3}+l_{4}}(t) & 
 =\frac{1+\mathfrak{\mathfrak{c}+\bar{\mathfrak{c}}}}{\bar{\mathfrak{c}}}(t)\frac{\bar{\mathfrak{c}}\mathfrak{B}}{1+\bar{\mathfrak{c}}\mathfrak{B}}(t)
\end{align*}
we perform the Trotter limit $N\rightarrow\infty$.
We drop terms that do not contribute in the limit $N\rightarrow\infty$. Furthermore we also have to drop the term $3N\ln\cos\lambda$ because
our $R$-matrix (\ref{eq:R Hubbard}) is not normalized as remarked after (\ref{eq:R
  Hubbard}). Instead we have
$
\check{R}\left(\lambda,0\right)\check{R}\left(0,\lambda\right)=\cos^6\lambda,
$
which is the term to be dropped. Therefore we find
\begin{align}
\ln\varLambda^{\text{QTM}}(\lambda) & =\left(\frac{1}{2z_{+}(x)}-\frac{z_{-}(x)}{2}-\frac{U}{4}\right)\beta\nonumber \\
 & \quad+ \left(k\circ\ln\frac{1+\mathfrak{\mathfrak{c}+\bar{\mathfrak{c}}}}{1+\bar{\mathfrak{c}}\mathfrak{B}}\right)\left(s\right)+\left(k\circ\ln\frac{ \left( 1+\mathfrak{c}+\bar{\mathfrak{c}}\right) \bar{\mathfrak{C}}} { \left( 1+\bar{\mathfrak{c}}\mathfrak{B}\right)\bar{\mathfrak{c}}} \right) \left(s-i\frac{U}{2}\right) \nonumber\\
 & \quad+\intop_{\mathcal{L}}\frac{\mathrm{d}t}{2\pi i}\left[\ln g\left(t-i\frac{U}{2}\right)\right]^{\prime}\ln\mathfrak{C}\left(t\right)
+\intop_{\mathcal{L}}\frac{\mathrm{d}t}{2\pi i}\left[\ln g\left(t\right)\right]^{\prime}\ln\frac{1+\mathfrak{\mathfrak{c}+\bar{\mathfrak{c}}}}{\bar{\mathfrak{c}}}\left(t\right).\label{eq:EW Hubbard}
\end{align}
Note that on the right-hand side $x$ and $s=s\left(z_{-}\left(x\right)\right)$ depend on $\lambda$
via (\ref{eq:Parametrisierung QTM}) and (\ref{eq:s Variable}) and
that $g(t)$ also depends on $x$ (\ref{eq:Def g(s)}).

Furthermore note that for $\lambda=0$ the expression for $\ln\varLambda^{\text{QTM}}(\lambda)$
simplifies to the well-known result \cite{JKS98}, which yields
the host's contribution to the thermodynamics of our impurity model
\begin{align}
f_{\text{h}} & =-\frac{1}{\beta}\ln\varLambda^{\text{QTM}}\left(0\right),\label{eq:host Hubbard}\\
\ln\varLambda^{\text{QTM}}\left(0\right) & =-\frac{\beta U}{4}+\intop_{\mathcal{L}}\frac{\mathrm{d}t}{2\pi i}\left[\ln z\left(t-i\frac{U}{2}\right)\right]^{\prime}\ln\mathfrak{C}(t)\nonumber \\
 & \quad+\intop_{\mathcal{L}}\frac{\mathrm{d}t}{2\pi i}\left[\ln z(t)\right]^{\prime}\ln\frac{1+\mathfrak{c}+\bar{\mathfrak{c}}}{\bar{\mathfrak{c}}}(t).\label{eq:fh Hubbard-1}
\end{align}
The impurity contribution to the total free energy $F=L\,f_h+f_i$ is given by
\begin{align}
f_{\text{i}} & =-\frac{1}{\beta}\ln\varLambda^{\text{QTM}}(\nu),\label{eq:fi hubbard}
\end{align}
Equations (\ref{eq:NLIE Hubbard Form 1}), (\ref{eq:EW Hubbard}) - (\ref{eq:fi hubbard}) completely
describe the thermodynamical properties of the Hubbard model with impurity.

For most situations the presented equations need to be treated numerically
which is beyond the scope of this paper and will be the topic of a separate
publication.

For the so-called half-filled case, however, we are able to perform the
low-temperature analysis in Appendix B. There we evaluate the $T^2$
contribution of the impurity to the total free energy. The result for the
eigenvalue of the impurity transfer matrix is
\begin{align}\ln\varLambda^{\text{QTM}}\left(\lambda\right)= & -\frac{U\beta}{4}+\beta\intop_{-1}^{1}\frac{\mathrm{d}t}{2\pi\text{i}}\left[\ln\frac{g\left(t\right)}{\bar{g}\left(t\right)}\right]^{\prime}\left(\kappa\ast f\right)(t)\nonumber\\
 & +\frac{U}{12\beta}\frac{1}{J_{0}\left({2\pi i}/U\right)+J_{2}\left({2\pi i}/U\right)}\sum_{n=0}^{\infty}J_{2n}\left(\frac{2\pi i}{U}\right)z_{-}^{-2n}\left(\lambda\right),\label{eq:Ausdruck}
\end{align}
where $z_{-}(\lambda)$ is defined in (\ref{eq:Parametrisierung}).
Here and in the
remainder of the paper, $\ast$ denotes the convolution of two functions $\left(f\ast
g\right)(x)=\frac1{2\pi}\int dy f(x-y)g(y)$.
Note that
$z_{-}(\lambda)\to\infty$ for $\lambda\to 0$ in which case only the $n=0$ term
in the series contributes. The third term in expression (\ref{eq:Ausdruck}) is
real. The functions $\kappa\left(t\right)$ and $f\left(t\right)$ are even
functions and thus $\left(\kappa\ast f\right)(t)$ is also even. Examination of
the ratio $\gamma(t):={g(t)}/{\bar{g}(t)}$ for
$z_{\pm}\left(\nu\right)\in\mathbb{R}$ shows that $\gamma(t)/\gamma(-t)$ is
unimodular, and hence $(\ln\gamma)'(t)+(\ln\gamma)'(-t)$ is purely imaginary.
This renders the second term on the r.h.s. of equation (\ref{eq:Ausdruck})
real.

We would like to note that it is possible to find alternative expressions for
$\ln\varLambda^{\text{QTM}}(\lambda)$. These alternative expressions are based
on the thermodynamic Bethe ansatz (TBA) and are given in Appendix C.

\section{\label{sec:level1}Conclusion}
We constructed a Hubbard model with integrable impurity and derived the Bethe
ansatz equations for the Hamiltonian. The impurity leads to an additional
phase factor in the first set of the nested Bethe ansatz equations of the
row-to-row transfer matrix resp.~the Hamiltonian.  For the finite temperature
properties, two sets of non-linear integral equations were derived: The
infinitely many thermodynamic Bethe ansatz equations and the finitely many
non-linear integral equations in the quantum transfer matrix approach. In both
cases the actual integral equations are unmodified by the impurity, i.e.\ they are identical to those of the homogeneous chain. The impurity appears in new
integral expressions of the thermodynamical potential which for the impurity
are different from those for the host.

Within the framework of TBA we used the string hypothesis for the new Bethe
ansatz equations of the Hamiltonian and followed the traditional procedure
\cite{Tak72}. Here, the derivation of the expression for the impurity
contribution to the thermodynamical potential was relatively
straightforward. The complexity of the infinitely many TBA equations, however,
constrains the practical use of this expression.

The derivation of the impurity contribution in terms of the finitely many
auxiliary functions that appear in the quantum transfer matrix approach
\cite{JKS98} was much more involved. It filled a substantial part of this
paper and constitutes the major result which allows for practical
calculations. As an example of such calculations we calculated the low-temperature asymptotics in the case of half filling. We are convinced that these results realize a significant
extension of the established knowledge of the Hubbard model \cite{EFGKK05}.

The Hubbard chain with integrable impurity is interesting in its own
right. In this present work, the foundation was laid for the investigation of the finite-temperature behavior by numerically solving the finitely many non-linear integral equations. A truncation like for the TBA equations presented in the Appendix is not necessary.
We can now evaluate the specific heat and entropy numerically. At high temperature $T$ the impurity spin will decouple from the host, but at low temperatures $T$ the impurity spin will be screened. The resulting Kondo physics will depend on the system parameters. Another application of the presented work are chains with more than one impurity site. The behavior of this type of impurities is special, since the scattering of particles on the two impurities at positions $j$ and $k$ does not depend on the difference $j-k$. This follows obviously from the construction of the commuting family of transfer matrices.

Beyond this, the integrable Hubbard chain with impurity also allows for new studies of the Anderson impurity
model. In a future publication we will present a suitable continuum limit of the Hubbard model
leading to a non-interacting host that interacts with an impurity with spin
and charge degrees of freedom. In this way we will derive a new set of
finitely many non-linear integral equations for the celebrated integrable
Anderson impurity model. 

\bibliography{Hubbard}

\section{\label{sec:level1}Appendix A - Expressions for the Hamiltonian}
We find 
\begin{align*}
R_{jk}\left(\nu,0\right)= & e^{2h\left(\nu\right)}+e^{h\left(\nu\right)}\left(\cos^{-1}\nu-e^{h\left(\nu\right)}\right)\sum_{\sigma=\uparrow,\downarrow}\left(n_{j\sigma}+n_{k\sigma}\right)\\
 & +e^{h\left(\nu\right)}\left(e^{h\left(\nu\right)}-\frac{2}{\cos\nu}+\frac{e^{h\left(\nu\right)}}{\cos^{2}\nu}\right)\left(n_{j\uparrow}n_{j\downarrow}+n_{k\uparrow}n_{k\downarrow}\right)\\
 & +\left(e^{2h\left(\nu\right)}+\cos^{-2}\nu\right)\left(n_{j\uparrow}n_{k\downarrow}+n_{j\downarrow}n_{k\uparrow}\right)\\
 & +\left(1+e^{2h\left(\nu\right)}\right)\sum_{\sigma=\uparrow,\downarrow}n_{j\sigma}n_{k\sigma}+\left(1-e^{2h\left(\nu\right)}\right)\left(1+\cos^{-2}\nu\right)\sum_{\sigma=\uparrow,\downarrow}n_{j\sigma}n_{k\uparrow}n_{k\downarrow}\\
 & +\left(\frac{4e^{h\left(\nu\right)}}{\cos\nu}-\frac{1+e^{2h\left(\nu\right)}}{\cos^{2}\nu}-1-e^{2h\left(\nu\right)}\right)n_{j\uparrow}n_{j\downarrow}\left(\sum_{\sigma=\uparrow,\downarrow}n_{k\sigma}-2n_{k\uparrow}n_{k\downarrow}\right)\\
 & +\tan\nu\sum_{\sigma=\uparrow,\downarrow}\left(c_{j\sigma}c_{k\sigma}^{\dagger}-e^{2h\left(\nu\right)}c_{j\sigma}^{\dagger}c_{k\sigma}\right)\\
 & +e^{h\left(\nu\right)}\tan\nu\left(e^{h\left(\nu\right)}-\cos^{-1}\nu\right)\left(c_{j\uparrow}^{\dagger}n_{j\downarrow}c_{k\uparrow}+n_{j\uparrow}c_{j\downarrow}^{\dagger}c_{k\downarrow}+c_{j\uparrow}^{\dagger}c_{k\uparrow}n_{k\downarrow}+c_{j\downarrow}^{\dagger}n_{k\uparrow}c_{k\downarrow}\right)\\
 & +\tan\nu\left(\frac{e^{h\left(\nu\right)}}{\cos\nu}-1\right)\left(c_{j\uparrow}n_{j\downarrow}c_{k\uparrow}^{\dagger}+n_{j\uparrow}c_{j\downarrow}c_{k\downarrow}^{\dagger}+c_{j\uparrow}c_{k\uparrow}^{\dagger}n_{k\downarrow}+c_{j\downarrow}n_{k\uparrow}c_{k\downarrow}^{\dagger}\right)\\
 & +\tan\nu\left(\frac{2e^{h\left(\nu\right)}}{\cos\nu}-1-e^{2h\left(\nu\right)}\right)\left(c_{j\uparrow}^{\dagger}n_{j\downarrow}c_{k\uparrow}n_{k\downarrow}+n_{j\uparrow}c_{j\downarrow}^{\dagger}n_{k\uparrow}c_{k\downarrow}-c_{j\uparrow}n_{j\downarrow}c_{k\uparrow}^{\dagger}n_{k\downarrow}\right.\\
 & \hphantom{+\tan\nu\left(\frac{2e^{h\left(\nu\right)}}{\cos\nu}-1-e^{2h\left(\nu\right)}\right)\left(\right.\,}\left.-n_{j\uparrow}c_{j\downarrow}n_{k\uparrow}c_{k\downarrow}^{\dagger}\right)\\
 & +\tan^{2}\nu\left(c_{j\uparrow}^{\dagger}c_{j\downarrow}c_{k\uparrow}c_{k\downarrow}^{\dagger}+c_{j\uparrow}c_{j\downarrow}^{\dagger}c_{k\uparrow}^{\dagger}c_{k\downarrow}-e^{2h\left(\nu\right)}\left(c_{j\uparrow}^{\dagger}c_{j\downarrow}^{\dagger}c_{k\uparrow}c_{k\downarrow}+c_{j\uparrow}c_{j\downarrow}c_{k\uparrow}^{\dagger}c_{k\downarrow}^{\dagger}\right)\right).
\end{align*}
\newpage
Note that $R_{jk}\left(0,\nu\right)$ has a similar form 
\begin{align*}
R_{jk}\left(0,\nu\right)= & e^{-2h\left(\nu\right)}+e^{-h\left(\nu\right)}\left(\cos^{-1}\nu-e^{-h\left(\nu\right)}\right)\sum_{\sigma=\uparrow,\downarrow}\left(n_{j\sigma}+n_{k\sigma}\right)\\
 & +e^{-h\left(\nu\right)}\left(e^{-h\left(\nu\right)}-\frac{2}{\cos\nu}+\frac{e^{-h\left(\nu\right)}}{\cos^{2}\nu}\right)\left(n_{j\uparrow}n_{j\downarrow}+n_{k\uparrow}n_{k\downarrow}\right)\\
 & +\left(e^{-2h\left(\nu\right)}+\cos^{-2}\nu\right)\left(n_{j\uparrow}n_{k\downarrow}+n_{j\downarrow}n_{k\uparrow}\right)\\
 & +\left(1+e^{-2h\left(\nu\right)}\right)\sum_{\sigma=\uparrow,\downarrow}n_{j\sigma}n_{k\sigma}+\left(1-e^{-2h\left(\nu\right)}\right)\left(1+\cos^{-2}\nu\right)\sum_{\sigma=\uparrow,\downarrow}n_{j\sigma}n_{k\uparrow}n_{k\downarrow}\\
 & +\left(\frac{4e^{-h\left(\nu\right)}}{\cos\nu}-\frac{1+e^{-2h\left(\nu\right)}}{\cos^{2}\nu}-1-e^{-2h\left(\nu\right)}\right)n_{j\uparrow}n_{j\downarrow}\left(\sum_{\sigma=\uparrow,\downarrow}n_{k\sigma}-2n_{k\uparrow}n_{k\downarrow}\right)\\
 & +\tan\nu\sum_{\sigma=\uparrow,\downarrow}\left(c_{j\sigma}^{\dagger}c_{k\sigma}-e^{-2h\left(\nu\right)}c_{j\sigma}c_{k\sigma}^{\dagger}\right)\\
 & +e^{-h\left(\nu\right)}\tan\nu\left(e^{-h\left(\nu\right)}-\cos^{-1}\nu\right)\left(c_{j\uparrow}n_{j\downarrow}c_{k\uparrow}^{\dagger}+n_{j\uparrow}c_{j\downarrow}c_{k\downarrow}^{\dagger}+c_{j\uparrow}c_{k\uparrow}^{\dagger}n_{k\downarrow}+c_{j\downarrow}n_{k\uparrow}c_{k\downarrow}^{\dagger}\right)\\
 & +\tan\nu\left(\frac{e^{-h\left(\nu\right)}}{\cos\nu}-1\right)\left(c_{j\uparrow}^{\dagger}n_{j\downarrow}c_{k\uparrow}+n_{j\uparrow}c_{j\downarrow}^{\dagger}c_{k\downarrow}+c_{j\uparrow}^{\dagger}c_{k\uparrow}n_{k\downarrow}+c_{j\downarrow}^{\dagger}n_{k\uparrow}c_{k\downarrow}\right)\\
 & +\tan\nu\left(1+e^{-2h\left(\nu\right)}-\frac{2e^{-h\left(\nu\right)}}{\cos\nu}\right)\left(c_{j\uparrow}^{\dagger}n_{j\downarrow}c_{k\uparrow}n_{k\downarrow}+n_{j\uparrow}c_{j\downarrow}^{\dagger}n_{k\uparrow}c_{k\downarrow}-c_{j\uparrow}n_{j\downarrow}c_{k\uparrow}^{\dagger}n_{k\downarrow}\right.\\
 & \hphantom{+\tan\nu\left(1+e^{-2h\left(\nu\right)}-\frac{2e^{-h\left(\nu\right)}}{\cos\nu}\right)\left(\right.\,}\left.-n_{j\uparrow}c_{j\downarrow}n_{k\uparrow}c_{k\downarrow}^{\dagger}\right)\\
 & +\tan^{2}\nu\left(c_{j\uparrow}^{\dagger}c_{j\downarrow}c_{k\uparrow}c_{k\downarrow}^{\dagger}+c_{j\uparrow}c_{j\downarrow}^{\dagger}c_{k\uparrow}^{\dagger}c_{k\downarrow}-e^{-2h\left(\nu\right)}\left(c_{j\uparrow}^{\dagger}c_{j\downarrow}^{\dagger}c_{k\uparrow}c_{k\downarrow}+c_{j\uparrow}c_{j\downarrow}c_{k\uparrow}^{\dagger}c_{k\downarrow}^{\dagger}\right)\right).
\end{align*}
These formulas can be used in equation (\ref{eq:h_imp}) and (\ref{eq:Hamiltonian}). The impurity part of the Hamiltonian has the form
\[
h_{\text{imp}}=\alpha_{0}\left(\nu,U\right)\left(h_{l,i}+h_{i,r}\right)+\alpha_{2}\left(\nu,U\right)h_{l,r}+\alpha_{1}\left(\nu,U\right)\left[\left(h_{l,i}+h_{i,r}\right),h_{l,r}\right],
\]
where the prefactors $\alpha_{j}\left(\nu,U\right)$, $j=0,1,2$ are quite bulky
expressions in terms of the model parameters. However, $\alpha_0$ and
$\alpha_2$ are real, and $\alpha_1$ is imaginary for real $\nu$. Hence,
$h_{\text{imp}}$ is hermitian.

\section{\label{sec:level1}Appendix B - Half filling Case for low temperatures}
In the following we analyze the behavior of our model in the case of half
filling and low temperatures. In this case
$\mathfrak{c}\left(t\right)$ and $\mathfrak{\bar{c}}\left(t\right)$ in
equations (\ref{eq:NLIE Hubbard Form 1}) are obviously small. Hence, terms
like $\ln\left(1+\mathfrak{c}\left(t\right)\right)$ or $\ln\left(1+\mathfrak{\bar{c}\left(t\right)}\right)$ may be dropped, but
not $\ln\bar{\mathfrak{c}}\left(t\right)$.

The equations simplify by use of the definitions
$a\left(t\right)=\mathfrak{b}\left(t-\text{i}\frac{U}{4}\right)$ and
$\bar{a}\left(t\right)=\mathfrak{b}^{-1}\left(t+\text{i}\frac{U}{4}\right)$ as
well as reducing the contour integrals to integrals on the real axis. Using
the Fourier transformation we find in $k$-space
\begin{align}
\mathfrak{a}\left(k\right) & =-\frac{e^{-\frac{U}{4}\left(\left|k\right|-2k\right)}}{1+e^{-\frac{U}{2}\left|k\right|}}\beta\mathfrak{f}\left(k\right)+\frac{e^{-\frac{U}{2}\left|k\right|}}{1+e^{-\frac{U}{2}\left|k\right|}}\mathfrak{A}\left(k\right)-\frac{e^{-\frac{U}{2}\left(\left|k\right|+k\right)}}{1+e^{-\frac{U}{2}\left|k\right|}}\bar{\mathfrak{A}}\left(k\right),\nonumber \\
\bar{\mathfrak{a}}\left(k\right) & =-\frac{e^{-\frac{U}{4}\left|k\right|}}{1+e^{-\frac{U}{2}\left|k\right|}}\beta\mathfrak{f}\left(k\right)+\frac{e^{-\frac{U}{2}\left|k\right|}}{1+e^{-\frac{U}{2}\left|k\right|}}\bar{\mathfrak{A}}\left(k\right)-\frac{e^{-\frac{U}{2}\left(\left|k\right|+k\right)}}{1+e^{-\frac{U}{2}\left|k\right|}}\mathfrak{A}\left(k\right),\label{eq:NL=000130E Appendix}
\end{align}
where $\mathfrak{a}\left(k\right)=\mathcal{F}\left\{ \ln a\right\}
\left(k\right)$, $\bar{\mathfrak{a}}\left(k\right)=\mathcal{F}\left\{
\ln\bar{a}\right\} \left(k\right)$,
$\mathfrak{A}\left(k\right)=\mathcal{F}\left\{ \ln\left(1+a\right)\right\}
\left(k\right)$, $\bar{\mathfrak{A}}\left(k\right)=\mathcal{F}\left\{
\ln\left(1+\bar{a}\right)\right\} \left(k\right)$ and
\begin{equation}
  \mathfrak{f}\left(k\right)={4\pi}\frac{J_{1}\left(k\right)}{k}=2\pi\left[J_{0}\left(k\right)+J_{2}\left(k\right)\right].
\end{equation}
Note
that $\mathfrak{f}\left(k\right)$ is the Fourier transform of
\[
f\left(t\right)=4\sqrt{1-t^{2}}\Theta\left(1-t^{2}\right)=\intop_{\mathbb{R}}\frac{\mathrm{d}k}{2\pi}e^{\text{i}kt}\mathfrak{f}\left(k\right).
\]
The Fourier transform of (\ref{eq:NL=000130E Appendix}) is
\begin{align}
  \ln a & =-\beta e\ast f +\kappa\ast\ln A-\kappa_-\ast\ln\bar{A},\nonumber \\
  \ln \bar{a} & =-\beta e\ast f +\kappa\ast\ln \bar{A}-\kappa_+\ast\ln A,\label{A1}
\end{align}
where the functions $e(x)$ and $\kappa(x)$ are defined by
\begin{equation}
  e(x)=\frac1U\frac1{\cosh\frac{2\pi x}U},\quad
\kappa(x)=\intop_{\mathbb{R}}\frac{\mathrm{d}k}{2\pi}e^{\text{i}kx}\frac{e^{-\frac{U}{2}\left|k\right|}}{1+e^{-\frac{U}{2}\left|k\right|}}
\label{A2}
\end{equation}
and $\kappa_\pm(x)=\kappa(x\pm i {U}/{2})$.

The driving term in (\ref{A1}) has exponential asymptotics
\begin{equation}
e\ast f(x)\simeq \frac{2}U\mathfrak{f}\left(\frac{2\pi
    i}U\right) \exp\left(-\frac{2\pi\left|x\right|}{U}\right),\qquad(|x|\to\infty).
\end{equation}
By use of the dilog-trick we obtain in the low-temperature limit $(\beta\to\infty)$
\begin{equation}
\intop_{0 (-\infty)}^{\infty (0)} \mathrm{d}x
\exp\left(-\frac{2\pi\left|x\right|}{U}\right)\ln\left(A(x)\bar{A}(x)\right)
\simeq\frac{\pi U^2}{24\beta\mathfrak{f}\left({2\pi
    i}/U\right)}.
\end{equation}

Performing the same analysis for the free energy (\ref{eq:EW Hubbard}) leads
to a significant simplification. Again, $\ln\bar{\mathfrak{c}}\left(t\right)$
must be treated carefully. All explicit terms simplify to the constant
$-\frac{U\beta}{4}$. Almost all integral expressions disappear. The only term
left comes from the last integral expression in equation (\ref{eq:EW Hubbard})
\[
\ln\varLambda^{\text{QTM}}\left(\lambda\right)=-\frac{U\beta}{4}-\intop_{-1}^{1}\frac{\mathrm{d}t}{2\pi\text{i}}\left[\ln\frac{g\left(t\right)}{\bar{g}\left(t\right)}\right]^{\prime}\left(\hat{K}_{1}\oblong\ln\mathfrak{B}\right)(t),
\]
where $\bar{g}\left(t\right)$ is the function $g\left(t\right)$ on the second
sheet. Rewriting this expression in the functions $a\left(t\right)$ and
$\bar{a}\left(t\right)$ yields
\begin{equation}
\ln\varLambda^{\text{QTM}}\left(\lambda\right)=-\frac{U\beta}{4}+\intop_{-1}^{1}\frac{\mathrm{d}t}{2\pi\text{i}}\left[\ln\frac{g\left(t\right)}{\bar{g}\left(t\right)}\right]^{\prime}
\left(\beta \left(\kappa\ast f\right)(t)+\left(e\ast\ln\left(A\bar A\right)\right)\left(t\right)\right).\label{A3}
\end{equation}
We want to evaluate the $\beta^{-1}$ contribution to this integral. It is
given by the $e\ast\ln\left(A\bar A\right)$ term which has low-temperature
asymptotics
\begin{equation}
\left(e\ast\ln\left(A\bar A\right)\right)\left(t\right)\simeq\frac{\pi U}{6\beta\mathfrak{f}\left({2\pi
    i}/U\right)}\cosh\frac{2\pi t}U.
\end{equation}

By use of the generating function of Bessel's functions we carry out the
integral and obtain (\ref{eq:Ausdruck}).

\section{\label{sec:level1}Appendix C - TBA}
In this appendix we apply alternatively to the main body of the paper the
thermodynamic Bethe ansatz (TBA). For the Hubbard model with impurity site we
will see that the non-linear integral equations \cite{Tak72} as obtained for
the pure Hubbard model (\ref{eq:H Hubbard}) still hold as well as the integral
expression for the host's contribution to the thermodynamical
potential. However, a new integral expression for the impurity's contribution
appears.

We use the string hypothesis \cite{Tak72,EFGKK05} according to which all
finite solutions of $\left\{ k_{j}\right\} _{j=1}^{K}$ and $\left\{
\varLambda_{l}\right\} _{l=1}^{M}$ of (\ref{eq:TM BAE Hubbard + imp}) are
composed of three different classes of strings
\begin{itemize}
\item a single real momentum $k_{j}$,
\item $m$ $\varLambda$'s combining into a $\varLambda$ string,
\item $2m$ $k$'s and $m$ $\varLambda$'s combining into a $k$-$\varLambda$
string. 
\end{itemize}
For large lattices ($L\gg1$) and a large number of electrons ($K\gg1$), nearly
for all strings the imaginary parts of the $k$'s and $\varLambda$'s are evenly
spaced.

Solving the equations (\ref{eq:TM BAE Hubbard + imp}) is simplified by use of
the string hypothesis. For arbitrary values of $K$ electrons and $M$ down
spins any solution of (\ref{eq:TM BAE Hubbard + imp}) is described by a
distribution of strings with $M_{n}$ $\varLambda$-strings
and $M_{n}^{\prime}$ $k$-$\varLambda$ strings of length $n$ ($n=1,2,\ldots$)
and $\mathcal{M}_{e}$ single $k_{j}$'s. The numbers $\mathcal{M}_{e}$, $M_{n}$
and $M_{n}^{\prime}$ are occupation numbers of the string
configuration and satisfy the sum rules 
\begin{align*}
M & =\sum_{n=1}^{\infty}n\left(M_{n}+M_{n}^{\prime}\right),\\
K & =\mathcal{M}_{e}+\sum_{n=1}^{\infty}2nM_{n}^{\prime}.
\end{align*}
Subjecting the string distribution to equations (\ref{eq:TM BAE Hubbard + imp})
and using 
\[
e^{i\hat{\delta}\left(k_{j}\right)}:=\e^{2h(\nu)}\frac{\e^{\i k_{j}}/{z_{+}(\nu)}+1}{z_{-}(\nu)-\e^{\i k_{j}}}
\]
we find for even $L$ in logarithmic form Takahashi's
equations for the purely real centers of the strings.
In the thermodynamical limit $L\rightarrow\infty,$ $\frac{K}{L}$
and $\frac{M}{L}$ fixed, solutions of Takahashi's equations 
should be expressed in terms of distributions of particles $\rho^{p}\left(k\right)$,
$\sigma_{n}^{p}\left(\varLambda\right)$, $\sigma_{n}^{\prime p}\left(\varLambda\right)$
and the appropriate $\rho^{h}\left(k\right),\sigma_{n}^{h}\left(\varLambda\right),\sigma_{n}^{\prime h}\left(\varLambda\right)$.
In the thermodynamic limit Takahashi's equations can be expressed as coupled integral
equations involving 
 the root densities \cite{Tak72,EFGKK05}
for particles and holes
\begin{align}
\rho^{p}\left(k\right)+\rho^{h}\left(k\right) & =\frac{1}{2\pi}+\frac{\hat{\Delta}\left(k\right)}{L}+\cos k\sum_{n=1}^{\infty}\intop_{-\infty}^{\infty}\mathrm{d}\varLambda a_{n}\left(\varLambda-\sin k\right)\left(\sigma_{n}^{\prime p}\left(\varLambda\right)+\sigma_{n}^{p}\left(\varLambda\right)\right),\nonumber \\
\sigma_{n}^{h}\left(\varLambda\right) & =-\sum_{m=1}^{\infty}\left.A_{nm}\ast\sigma_{m}^{p}\right|_{\varLambda}+\intop_{-\pi}^{\pi}\mathrm{d}ka_{n}\left(\varLambda-\sin k\right)\rho^{p}\left(k\right),\nonumber \\
\sigma_{n}^{\prime h}\left(\varLambda\right) & =\frac{1}{\pi}\text{Re}\frac{1}{\sqrt{1-\left(\varLambda-ni\frac{U}{4}\right)^{2}}}+\frac{2}{\pi L}\partial_{\varLambda}\hat{\delta}\left(\text{Re}\sqrt{1-\left(\varLambda-ni\frac{U}{4}\right)^{2}}\right)\nonumber \\
 & \quad-\sum_{m=1}^{\infty}\left.A_{nm}\ast\sigma_{m}^{\prime p}\right|_{\varLambda}-\intop_{-\pi}^{\pi}\mathrm{d}ka_{n}\left(\sin k-\varLambda\right)\rho^{p}\left(k\right),\label{eq:TBA Dichten Hubbard}
\end{align}
where 
$
\hat{\Delta}\left(k\right)=\frac{1}{2\pi}\partial_{k}\hat{\delta}\left(k\right),
$
$a_{n}\left(x\right)$ is a shorthand notation for
\[
a_{n}\left(x\right)=\frac{1}{2\pi}\frac{n\frac{U}{2}}{\left(n\frac{U}{4}\right)^{2}+x^{2}},
\]
and $A_{nm}*$ is an integral operator that acts on a function $f\left(x\right)$ as
\[
\left.A_{nm}\ast f\right|_{x}=\delta_{nm}f\left(x\right)+\intop_{-\infty}^{\infty}\frac{\mathrm{d}y}{2\pi}\frac{\mathrm{d}}{\mathrm{d}x}\varTheta_{nm}\left(\frac{x-y}{U/4}\right)f\left(y\right).
\]
Note that in general $\ast$ denotes the convolution of two functions.
The ad hoc definition of $A_{nm}*$
avoids the introduction of a {\it function} $A_{nm}$ with delta function contribution.

For further transformations of (\ref{eq:TBA Dichten Hubbard}) the
following relation is of great use
\[
\intop_{-\pi}^{\pi}\frac{\mathrm{d}k}{2}a_{n}\left(\varLambda-\sin k\right)=\text{Re}\frac{1}{\sqrt{1-\left(\varLambda-ni\frac{U}{4}\right)^{2}}}.
\]
To find the state of thermodynamic equilibrium 
we locate, following the principles of statistical mechanics,
the minimum of the thermodynamical
potential \cite{TW83} per site
\[
f=e-\mu n_{c}-2Bm-Ts.
\]
where
$\mu$ is the chemical potential, $B$ the magnetic field, $T$
the temperature, $n_{c}$ the particle density, $m$ the magnetization
and $s$ is the total entropy per site. We restrict ourselves to a magnetic
field $B\geq0$ and to a chemical potential $\mu\leq0$.

In the thermodynamic limit, we use the root densities of particles and holes
and consider the entropy as a functional in terms of the root densities.

We use Stirling's formula to approximate the factorials in the contribution
$\mathrm{d}S$ to the entropy, since the logarithm of the number
of states is large in the thermodynamic limit.
The thermodynamical potential per site $f$ is a functional in terms of the
root densities. With respect to variations in a maximal set of independent
root densities the state of thermodynamic equilibrium must be a stationary point. Equations (\ref{eq:TBA Dichten Hubbard}) give the
densities of holes in terms of the densities of particles. Hence the
variational condition $\delta f=0$
is to be solved under the constraint equations (\ref{eq:TBA Dichten
  Hubbard}). This results into the thermodynamical Bethe ansatz equations just
for the ratios 
\[
\zeta\left(k\right)=\frac{\rho^{h}\left(k\right)}{\rho^{p}\left(k\right)},\qquad\eta_{n}\left(\varLambda\right)=\frac{\sigma_{n}^{h}\left(\varLambda\right)}{\sigma_{n}^{p}\left(\varLambda\right)},\qquad\eta_{n}^{\prime}\left(\varLambda\right)=\frac{\sigma_{n}^{\prime h}\left(\varLambda\right)}{\sigma_{n}^{\prime p}\left(\varLambda\right)},
\]
and as mentioned already above are identical to those of
the pure Hubbard model \cite{Tak72}
\begin{align}
\ln\zeta\left(k\right) & =-\frac{2\cos
  k}{T}-\frac{4}{T}\intop_{-\infty}^{\infty}\mathrm{d}y\, s(\sin k-y)\text{Re}\sqrt{1-\left(y-i\frac{U}{4}\right)^{2}}
+\left(s\ast\ln\frac{1+\eta_{1}^{\prime}}{1+\eta_{1}}\right)\left(\sin k\right),\nonumber \\
\eta_{0}\left(\varLambda\right) & =\eta_{0}^{\prime}\left(\varLambda\right)=0,\nonumber \\
\ln\eta_{n}\left(\varLambda\right) & =\left(s\ast\ln\left(\left(1+\eta_{n-1}\right)\left(1+\eta_{n+1}\right)\right)\right)\left(\varLambda\right)-\delta_{1n}\left(s\ast\ln\left(1+\zeta^{-1}\right)\right)\left(\varLambda\right),\nonumber \\
\ln\eta_{n}^{\prime}\left(\varLambda\right) & =\left(s\ast\ln\left(\left(1+\eta_{n-1}\right)\left(1+\eta_{n+1}\right)\right)\right)\left(\varLambda\right)-\delta_{1n}\left(s\ast\ln\left(1+\zeta\right)\right)\left(\varLambda\right),\label{eq:TBA Hubbard}
\end{align}
for $n=1,2,\ldots$ and with integral kernel $
s\left(x\right)=\frac{1}{U\text{ch}\frac{2\pi x}{U}}$ in the convolutions.
Equations (\ref{eq:TBA Hubbard}) are completed by the boundary conditions
\[
\lim_{n\rightarrow\infty}\frac{\ln\eta_{n}}{n}=\frac{2B}{T},\qquad\lim_{n\rightarrow\infty}\frac{\ln\eta_{n}^{\prime}}{n}=-\frac{2\mu}{T}.
\]
The thermodynamical potential per site of the host is given in terms of solutions
to (\ref{eq:TBA Hubbard}) as
\begin{align}
f_{\text{h}} & =\frac{U}{4}-T\intop_{-\pi}^{\pi}\frac{\mathrm{d}k}{2\pi}\ln\left(1+\zeta^{-1}\left(k\right)\right)\nonumber \\
 & \quad-T\sum_{n=1}^{\infty}\intop_{-\infty}^{\infty}\frac{\mathrm{d}\varLambda}{\pi}\ln\left(1+\left(\eta_{n}^{\prime}\right)^{-1}\left(\varLambda\right)\right)\text{Re}\frac{1}{\sqrt{1-\left(\varLambda-in\frac{U}{4}\right)^{2}}}.\label{eq:fh Hubbard}
\end{align}
The total thermodynamical potential is of the form
\begin{equation}
F=Lf_{\text{h}}+f_{\text{i}},\label{eq:f Hubbard}
\end{equation}
where $f_{\text{i}}$ is the impurity part of the thermodynamical
potential 
\begin{align}
f_{\text{i}} & =\frac{U}{4}-T\intop_{-\pi}^{\pi}\mathrm{d}k\hat{\Delta}\left(k\right)\ln\left(1+\zeta^{-1} \left(k\right)\right) \nonumber \\
 & \quad-T\sum_{n=1}^{\infty}\intop_{-\pi}^{\pi}\mathrm{d}k\intop_{-\infty}^{\infty}\mathrm{d}\varLambda\hat{\Delta}\left(k\right)a_{n}\left(\varLambda-\sin k\right)\ln\left(1+\left( \eta_{n}^{\prime}\right)^{-1} \left(\varLambda\right)\right)
.\label{eq:fi-Hubbard}
\end{align}

Equations (\ref{eq:TBA Hubbard}), (\ref{eq:fh Hubbard}), (\ref{eq:fi-Hubbard})
and (\ref{eq:f Hubbard}) completely describe the thermodynamical properties of
the Hubbard model with impurity. The equivalence of the thermodynamical
equations was shown in \cite{CCMT15}. Note that for
$z_{\pm}\left(\nu\right)\in\mathbb{R}$ (that is, for example, for
$\nu\in\mathbb{R}$), expression (\ref{eq:fi-Hubbard}) yields a real result:
The odd part $\hat{\delta}\left(k\right)-\hat{\delta}\left(-k\right)$ of
the function $\hat{\delta}\left(k\right)$ is real, hence
$\hat{\Delta}\left(k\right)+\hat{\Delta}\left(-k\right)$ is real too and all
other factors in (\ref{eq:fi-Hubbard}) are even in $k$. Therefore the result of
(\ref{eq:fi-Hubbard}) is real.
\section{\label{sec:level1}Acknowledgment}
YÖ is grateful to Studienstiftung des deutschen Volkes (German Academic 
Scholarship Foundation) for a PhD grant. Both authors acknowledge financial 
support by Deutsche Forschungsgemeinschaft within FOR 2316 "Correlations in 
Integrable Quantum Many-Body Systems".
\end{document}